\documentclass[twocolumn,epjc3]{svjour3}

\DeclareSymbolFont{matha}{OML}{txmi}{m}{it}% txfonts
\DeclareMathSymbol{\varv}{\mathord}{matha}{118}
\usepackage{amssymb}
\usepackage{xcolor}
\usepackage{soul,ulem}
\usepackage{physics}
\usepackage{upgreek,textgreek}
\smartqed  % flush right qed marks, e.g. at end of proof
\RequirePackage{graphicx}

\RequirePackage{latexsym}
\RequirePackage[numbers,sort&compress]{natbib}
\RequirePackage[colorlinks,citecolor=blue,urlcolor=blue,linkcolor=blue]{hyperref}

\usepackage{commath} % \abs{} -> |a|; \norm{] -> ||a||
	\usepackage{braket}  % \braket{} -> <a>
	\usepackage{lipsum} 
	\usepackage{import}
	\usepackage{color}
\usepackage{bm}% bold math
\usepackage{dcolumn}% Align table columns on decimal point
\usepackage[utf8]{inputenc}
\usepackage[absolute,overlay]{textpos}
\usepackage{hyperref}
\usepackage{graphicx}
\usepackage{dcolumn}
\usepackage{xcolor}
\usepackage[numbers,sort&compress]{natbib}
\usepackage[all]{nowidow}
\let\oldAA\AA
\renewcommand{\AA}{\text{\normalfont\oldAA}}

\journalname{Eur. Phys. J. Plus}
\begin{document}

\title{Many-Body Dissipative Particle Dynamics with the MARTINI ``Lego'' Approach%\thanksref{t1}
}
% \subtitle{Do you have a subtitle?\\ If so, write it here}


% \author{First Author\thanksref{e1,addr1}
%         \and
%         Second Author\thanksref{e2,addr2,addr3} %etc.
% }

\author{Lu\'is H. Carnevale\thanksref{e1,addr1}
        \and
        Panagiotis E. Theodorakis\thanksref{e2,addr1} %etc.
}

\thankstext{e1}{e-mail: carnevale@ifpan.edu.pl}
\thankstext{e2}{e-mail: panos@ifpan.edu.pl}


% \institute{First address \label{addr1}
%            \and
%            Second address \label{addr2}
%            \and
%            \emph{Present Address:} if needed\label{addr3}
% }
\institute{Institute of Physics, Polish Academy of Sciences, Al. Lotnik\'ow 32/46, 02-668 Warsaw, Poland \label{addr1}
}

\date{Received: date / Accepted: date}

\maketitle

\begin{abstract}
MARTINI is a popular coarse-grained (CG) force-field that is used in
molecular dynamics (MD) simulations. It is based on the ``Lego'' approach
where nonbonded interactions between CG beads 
representing chemical units of different polarity are obtained
through water--octanol partition coefficients. This enables the
simulation of a wide range of molecules by only using a finite number of
parametrized CG beads, similar to the Lego game, 
where a finite number
of brick types is used to create larger structures.
Moreover, the MARTINI force-field is based on the Lennard-Jones potential with
the shortest possible cutoff including attractions,
thus rendering it very efficient for MD simulations. 
However, MD simulation is in general
a computationally expensive method. Here, we demonstrate that using the
MARTINI ``Lego'' approach is suitable for many-body dissipative
particle (MDPD) dynamics, a method that can simulate multi-component 
and multi-phase soft matter systems in a much faster time
than MD. In this study, a DPPC lipid
bilayer is chosen to provide evidence for the validity of this
approach and various properties are compared to highlight the potential
of the method, which can be further extended by introducing new CG bead types. 
\end{abstract}

\section{Introduction}
Molecular dynamics \cite{Rapaport_2004} (MD) has
been established as an indispensable tool
for scientific research in a wide range of areas, 
such as solid state physics, soft matter, fluid dynamics, 
and biophysics.
The method has also become popular due to the 
existence of well-documented, open-source, free,
parallel software, such as LAMMPS \cite{Plimpton1995,Thompson2022} 
and GROMACS \cite{Berendsen1995}, which are used and
supported by vast and vibrant scientific communities. 
In the heart of a successful MD simulation lies the
force-field (model), which describes the interactions between
the particles (or group of particles/atoms) in a system.
The force-field should be able to faithfully reproduce
specific properties (\textit{e.g.}, surface tension) of the system
under specific conditions (\textit{e.g.}, a certain temperature or pressure).
For example, in the case of systems with surfactants,
the viscous and surface-tension forces are important
properties that the force-field should capture well,
in order to describe phenomena, such as droplet coalescence \cite{Arbabi2023}.
Therefore, it is natural that much of research in MD 
(MD force-fields are also directly suitable for the Monte Carlo method \cite{landau_binder_2014})
concerns the development of high-quality force-fields
that renders a simulation reliable to reproduce system's
properties.

However, MD simulations are generally computationally 
expensive, especially in the case of all-atom force-fields. 
This has led to the quest for more computationally 
efficient force-fields, where a single interaction site
would correspond to a group of atoms, namely coarse-grained (CG) models \cite{Noid2013,Kmiecik2016,Jin2022,Noid2023}.
In the case of CG force-fields the number
of interacting sites that represent the system is therefore reduced,
which translates into less computations,
and in turn importantly a much smaller carbon footprint of simulations.
To this end, one of the most popular CG force-fields is
MARTINI \cite{Marrink2004,Marrink2007,Marrink2013,Alessandri2021},
which has been proposed almost two decades ago
and is still under development with the MARTINI 3.0 \cite{Souza2021}
recently released. The MARTINI force-field can be used to simulate
a wide spectrum of systems, such as proteins \cite{Monticelli2008,Periole2009,deJong2013,Poma2017},
polymers \cite{Lee2009}, carbohydrates \cite{Lopez2009},
glycolipids \cite{Lopez2013}, glycans \cite{Chakraborty2021},
DNA \cite{Uusitalo2015}, RNA \cite{Uusitalo2017}, water \cite{Yesylevskyy2010}
and various solvents \cite{Vainikka2021}, while various extensions include
simulations for specific pH \cite{Grunewald2020}
and more recently chemical reactions
(reactive MARTINI) \cite{Sami2023}.
Without going into the nitty-gritty details of the
MARTINI force-field here (some of them will be discussed
in the next Section), the reasons for its popularity
are manifold. Most importantly, it is versatile and transferable, thus
enabling the simulation of a wide range of systems, as
mentioned above. This is mainly due to the ``Lego'' approach,
where beads with different polarity can be used to
represent various chemical units, which in turn are
used to build the various molecules of the system,
that is complex soft matter systems of different components. 
Such an approach might hinder a detailed representation of
a chemical unit by a CG bead, but this might not always be
required for the CG modeling of a system.
MARTINI also uses the shortest possible cutoff 
of Lennard-Jones interactions including attractions, 
which renders MARTINI computationally efficient for MD. 
Moreover, it is actively developed and tested by
a large community with information, examples and applications
being abundant in the literature and online.

MARTINI has therefore been an established force-field 
for MD simulation. However, another computational method beyond MD
could probably offer further possibilities for faster simulations. 
Importantly, this method should include attractive interactions in order
to enable the simulation of multiphase systems.
Here, we show that the many-body dissipative particle 
dynamics (MDPD) \cite{Pagonabarraga2001,warren2003,zhao2017,zhao2021-review,zhao2021,Han2021} 
method can be used to substitute MD 
simulations in certain areas, and, when combined with the MARTINI
``Lego'' approach, can provide accurate descriptions
of a system at a much lower computational cost. 
As in the case of the MARTINI approach, the intermolecular
interactions between CG beads are obtained via a top--down
approach based on the water--octanol partition coefficients \cite{Anderson2017}.
In general, MDPD is suitable for simulating multi-phase and
multi-component systems that are much larger than in
MD \cite{Carnevale2023,Carnevale2024,zhao2021-review}.
Moreover, in view of the rapid development of MDPD
with applications in soft matter and fluids physics, we anticipate that our study
opens new possibilities for this method in these areas
by taking advantage of the MARTINI force-field approach.

In the next section,
our methods and system descriptions are provided. 
Then, we juxtapose the results
of the MD MARTINI with the MDPD MARTINI model for a
lipid system, demonstrating the reliability of the MDPD 
MARTINI model with an accompanied lower computational cost. 
Then, in the last section,
we draw our conclusions and outline the prospects
for further development.

\section{Theoretical Methods}\label{methods}

\subsection{Molecular Dynamics}
We first simulate the formation of a DPPC lipid bilayer
in water with MD simulation. 
This system has been used as a benchmark in
the MARTINI online tutorials \cite{lipids_tutorial} 
by using GROMACS software \cite{Berendsen1995}.
Here, we have used LAMMPS \cite{Plimpton1995,Thompson2022} software to carry out both
the MD and the MDPD simulations. Regarding the MD simulations,
initial files for the simulation of the DPPC bilayer can
be found as a moltemplate example \cite{lipids_tutorial_moltemplate}.
In particular, we initiated simulations for systems comprising
128, 256, and 512 lipids. These lipid-containing systems were 
initially placed within a simulation box characterized by
dimensions $L_x=L_y=L_z=70$, $90$, and $120$, respectively, 
in conjunction with a water component maintaining a 10:1 proportion
of water beads. The MARTINI DPPC model, employed in our simulations,
encompasses a representation consisting of distinct beads:
$Q_0$ to signify the choline headgroup, $Q_a$ beads representing
the phosphate headgroup, two $N_a$ beads embodying 
the glycerol ester moiety, and two chains, 
each comprised of four $C_1$ beads, to emulate the alkane tails. 
Water entities were simulated using $P_4$ beads.

To initiate our simulations, we adapted the example files provided
by moltemplate \cite{lipids_tutorial_moltemplate}.
We commenced the simulations by initially minimizing the system's 
energy, followed by conducting a simulation in the NPT ensemble
employing the Nosé--Hoover thermostat at 300K and 1 bar.
This step aimed to relax the system and attain an equilibrium box
volume. Subsequently, we executed simulations under NVT conditions,
elevating the temperature to 450K to facilitate the removal of any
undesired structures that may have formed, paving the way for the
bilayer formation. Further refinement of the system was achieved by running simulations
within the NPT ensemble, employing a zero surface tension condition
while gradually lowering the temperature to 300~K. 
This phase culminated in the attainment of the final bilayer 
configuration, characterized by the absence of defects. 
Lastly, we kept the NPT run at 300K to systematically measure 
key bilayer properties.

\subsection{Many-body dissipative particle dynamics}
The MDPD method has been applied
for fluids with different properties \cite{Espanol1995,warren2003,zhao2017,zhao2021-review,zhao2021,Han2021,Vanya2018},
such as density and surface tension \cite{Carnevale2023},
as well as a range of soft matter multi-component
systems, such as systems with surfactant molecules \cite{Hendrikse2023}.
In the case of the MDPD model, the Langevin equation (Equation~\ref{eq1}) of
motion is solved as is done in MD Langevin dynamics \cite{Theodorakis2011},
but forces are directly defined in the MDPD method rather than
derived from a potential form as in the case of MD.
Equation~\ref{eq1} describes the motion of each particle, $i$,
interacting with its neighbors, $j$, through a conservative
force $\bm{F}^C$, namely,
\begin{eqnarray}
	m\frac{d\bm{v}_i}{dt} = \sum_{j\neq i} \bm{F}_{ij}^C + \bm{F}_{ij}^R + \bm{F}_{ij}^D.
	\label{eq1}
\end{eqnarray}

The random force, $\bm{F}^R$, and the dissipative force, $\bm{F}^D$,
act on each particle, $i$, and function as system's thermostat, since
they are related via the fluctuation--dissipation theorem.
The mass $m$ is kept the same for all particles and set to unity.
A main difference between the MDPD and the 
DPD \cite{Espanol2017,Yoshimoto2013,Li2014,Lavagnini2021,Trofimov2005} approaches 
is the expression of the conservative force, which also includes
attractive interactions, and reads
\begin{eqnarray}
	\bm{F}^C_{ij} =  A\omega^C(r_{ij})\bm{e}_{ij} + 
	B \left(\bar{\rho_i} + \bar{\rho_j} \right) \omega^d(r_{ij})\bm{e}_{ij}.
	\label{eq2}
\end{eqnarray}
$A<0$ and $B>0$ are parameters of the force-field, which
correspond to attractive and repulsive interactions, respectively.
$r_{ij}$ is the distance between particles, 
$\bm{e}_{ij}$ the unit vector connecting particles $i$ and $j$,
while $\omega^C(r_{ij})$ and $\omega^d(r_{ij})$ are linear
weight functions defined as
\begin{eqnarray}
	\omega^{C}(r_{ij}) = 
	\begin{cases}
		&1 - \frac{r_{ij}}{r_{c}}, \ \ r_{ij} \leq r_{c} \\
		& 0,  \  \ r_{ij} > r_{c}.
	\end{cases} 
	\label{eq3}
\end{eqnarray}
Here, $r_c$ is a cutoff distance for the interactions, 
which is usually set to unity. Moreover, $\omega^d(r_{ij})$ is of the same form,
however, its cutoff distance $r_d=0.75$, smaller than $r_c$.

The many-body contributions to the force-field are included
in the repulsive term and depends on the local neighborhood densities,
$\bar{\rho_i}$ and $\bar{\rho_j}$, via the following expression:
\begin{eqnarray}
	\bar{\rho_i} = 
	\sum_{0<r_{ij}\le r_d}
	\frac{15}{2\pi r_d^3} \left( 1 - \frac{r_{ij}}{r_d}\right)^2 .
	\label{eq4}
\end{eqnarray}
The random and dissipative forces read
\begin{eqnarray}
	\bm{F}^D_{ij} = -\gamma \omega^D(r_{ij}) (\bm{e}_{ij} \cdot  \bm{v}_{ij})\bm{e}_{ij} ,
	\label{eq5}
\end{eqnarray}
\begin{eqnarray}
	\bm{F}^R_{ij} = \xi \omega^R(r_{ij}) \theta_{ij} \Delta t^{-1/2} \bm{e}_{ij} ,
	\label{eq26}
\end{eqnarray}
with $\gamma$ being the dissipative strength, 
$\xi$ the strength of the random force, 
$\bm{v}_{ij}$ the relative velocity between particles, and $\theta_{ij}$ a random 
variable from a Gaussian distribution with unit variance.
In this case, the fluctuation--dissipation theorem require that
$\gamma$ and $\xi$ are related to each other by
\begin{eqnarray}
	\gamma = \frac{\xi ^2}{2 k_B T}.
	\label{eq7}
\end{eqnarray}
The temperature of the system in our simulations is set to $T=1$ (MDPD units)
while the weight functions for the forces are
\begin{eqnarray}
	\omega^D(r_{ij}) = \left[\omega^R(r_{ij})\right]^2 = 
	\begin{cases}
		\left( 1 - \frac{r_{ij}}{r_c}\right)^2,  & r_{ij} \leq r_{c} \\
		0,   & r_{ij} > r_{c}.
	\end{cases}
	\label{eq8}
\end{eqnarray}
Finally, the equations of motion (Equation~\ref{eq1})
are integrated by using a modified velocity-Verlet algorithm as
implemented in LAMMPS software \cite{Plimpton1995,Thompson2022}
with a time step $\Delta t = 0.01$.
Describing the various interactions for the system
amounts to determining the attractive parameter, $A_{ij}$, of
the conservative force between pair of particles, while
the repulsive part of the interactions expressed through $B$
will remain constant \cite{warren2013}. 
To build molecules, one also needs to use bond and
angle potentials. In particular,
harmonic interactions, mathematically expressed as
\begin{equation}
    E_{bond} = \frac{k}{2} \left( r_{ij} - r_0 \right)^2
\end{equation}
are used to bind consecutive beads in order to be able to form
a macromolecular chain (\textit{e.g.}, lipid, surfactant).
Here, as is usually done in the case of CG models \cite{muller2014force}
and for the sake of simplicity,
the same parameters are used for all bonds,
namely $k = 150$ and $r_0 = 0.5$.
Moreover, a harmonic angle potential is applied for
a triad of sequential beads along the molecular chains,
\begin{equation}
    E_{angle} = \frac{k_A}{2} \left( \theta_{ijk} - \theta_0 \right)^2, 
\end{equation}
where $k_A = 5$ and $\theta_0$, which depends on
the specific molecule. In the following section, we discuss the 
force-field parametrization of the intermolecular interactions.

\subsection{Force-field parametrization}
To obtain the water--octanol partition coefficients,
we have measured the molar concentrations of the solute
molecules in the water, $[S]_{w}$, and 1-octanol phases, $[S]_{o}$, 
following a previous study \cite{Anderson2017}.
Then, the water--octanol partition coefficient is
defined as
\begin{eqnarray}
\label{Eq:owpart}
	\log(P_{o/w}) = \log\frac{[S]_{o}}{[S]_{w}}.
\end{eqnarray}
The advantage of this approach based on the
water--octanol partition coefficient lies in the fact that
experimental data are abundant for various solute
molecules \cite{Sangster1989,Sangster1997}. 
Moreover, the coefficient can be readily calculated
in particle-based simulations by directly using
the above equation and determining the quantities
$[S]_{w}$ and $[S]_{o}$, or via umbrella sampling
simulations, as has previously been done in the
case of the MARTINI model \cite{Marrink2013,Souza2021}. 
However, umbrella sampling simulations are not
expected to provide any advantage in the case
of MDPD simulations, since MDPD uses soft interactions
that provide better sampling and a more efficient path
to equilibrium. Moreover, umbrella simulations would require
an expression for the underlying potential, which 
may further complicate things. Here, a limitation in measuring
$\log P$ with the current approach comes from the
requirement that at least one bead is present in each 
phase. In this case, the maximum $\log P$ value that
can be obtained is $\log N$, when the number of
solute beads is $N$. Our system contains 1500
solute beads (around 3\% of the concentration),
which implies a limit $\log(P_{o/w}) = \pm 3.17$
able to describe both apolar and polar solutes.
This is in general satisfactory, since
the usual range in experiments is between $-4$ and $7$.

Here, we have chosen to parametrize a composite system consisting
of water and DPPC lipids \cite{Zhu2021}, which serves
as a benchmark in the simulation of lipid membranes.
According to the
MARTINI model, four heavy atoms are usually represented by
one CG bead. While we follow here the MARTINI recipe, 
a different CG level might in general be chosen, for example as
in the case of models based on the Statistical Associating Fluid Theory \cite{muller2014force}. 
The parametrization process outlined in this study involves several crucial 
steps to characterize and define the interactions between different types
of beads in a complex molecular system.

\begin{figure}[bt!]
\centering
\includegraphics[width=\columnwidth]{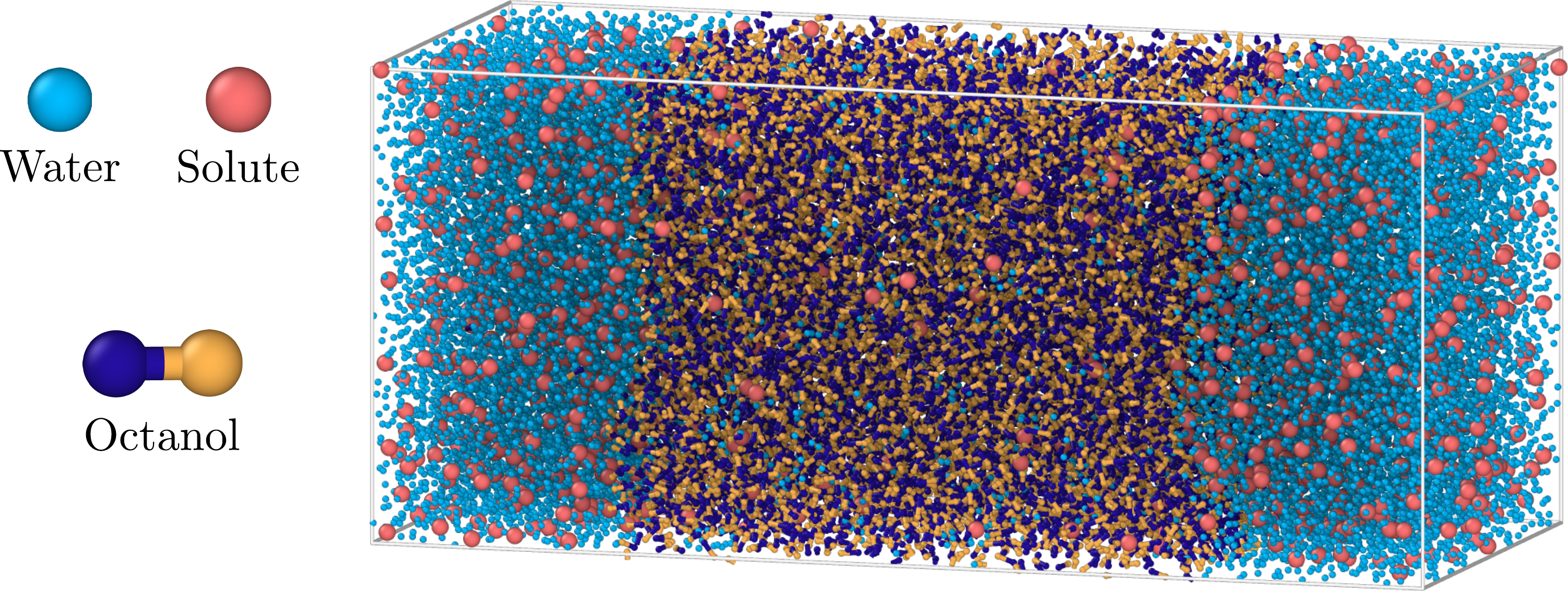}
\caption{\label{fig:1} 
(a) Typical simulation snapshot of a water--octanol system
used for the parametrization of a solute particle (H bead) 
based on the o/w partitioning (Equation~\ref{Eq:owpart}). 
}
\end{figure}

Firstly, the system is represented by three primary bead types 
in the reference system, namely the water (W) bead, the octanol (C) bead
used for the alkane chain, and the G bead representing
the polar alcohol portion of the chain. Notably, the W bead corresponds to
three water molecules, while the C bead is employed to model a four carbon
apolar chain, and the G bead captures the polar components.

Secondly, the self-interaction parameters, denoted as $A_{ii}$, 
are determined for each bead type. 
These parameters are derived from surface tension measurements 
and are specific to different bead types. 
For instance, $A_{WW}$ pertains to water, $A_{CC}$ corresponds 
to octane (comprised of two bonded C beads), and $A_{GG}$ is associated 
with Diethylene Glycol \cite{Hendrikse2023} (composed of two bonded G beads).
The surface tension can be measured by placing a slab of the desired 
component in a large simulation box and as usual computed from the pressure 
tensor (mechanical approach \cite{Kirkwood1949}) as
\begin{equation}
    \gamma = \frac{L_z}{2}\left( P_{zz} - \frac{P_{xx} + P_{yy}}{2} \right),
\end{equation}
where $P_{zz}$ is the pressure in the direction normal to the surface.
The cross-interaction parameter between C and G beads, specifically $A_{CG}$, is also calculated based on surface tension measurements involving octanol.

Thirdly, additional cross-interaction parameters,
namely $A_{WC}$ and $A_{WG}$, are determined through a comparison 
with the partition coefficient ($\log(P_{o/w})$) for water beads 
in the octanol phase and octanol molecules in the water phase. 
This comparison aligns with experimental data, with water in octanol
exhibiting a $\log(P)$ of 1.3 (experimental) and 1.4 (simulated), 
while octanol in water shows a $\log(P)$ of 3.1 (experimental)
and 2.9 (simulated). The accuracy of the model is further validated
by comparing the calculated Diethylene Glycol $\log(P)$ values 
(-1.5), which closely match the experimental value -1.4.

Lastly, the parametrization process extends to other bead types 
by categorizing them as either polar or apolar based on their 
behavior and measuring their partition coefficients ($\log(P)$) 
to capture their respective characteristics.
An illustrative example is provided in Figure \ref{fig:1}, 
where the H bead (solute) in the lipid model is identified 
as highly polar with a $\log(P)$ value of -1.7. All the 
attractive parameters are presented in the interaction matrix 
of Table \ref{tab:1}.

\begin{table}[b]
\begin{tabular}{c|cccc}
    & W  & H  & G   & C \\
  \hline
  W & I  & I  & II  & V \\ 
  H & I  & IV & II  & V \\ 
  G & II & II & III & V \\ 
  C & V  & V  &  V  & V \\ 
\end{tabular}
\caption{Interaction matrix for parametrized MDPD beads organized in five interaction levels (I-V) with corresponding attractive parameters (A):
$-50$ (I); $-43$ (II); $-34$ (III); $-30$ (IV); $-26$ (V). }
\label{tab:1} 
\end{table}

\begin{figure}[bt!]
\centering
\includegraphics[width=\columnwidth]{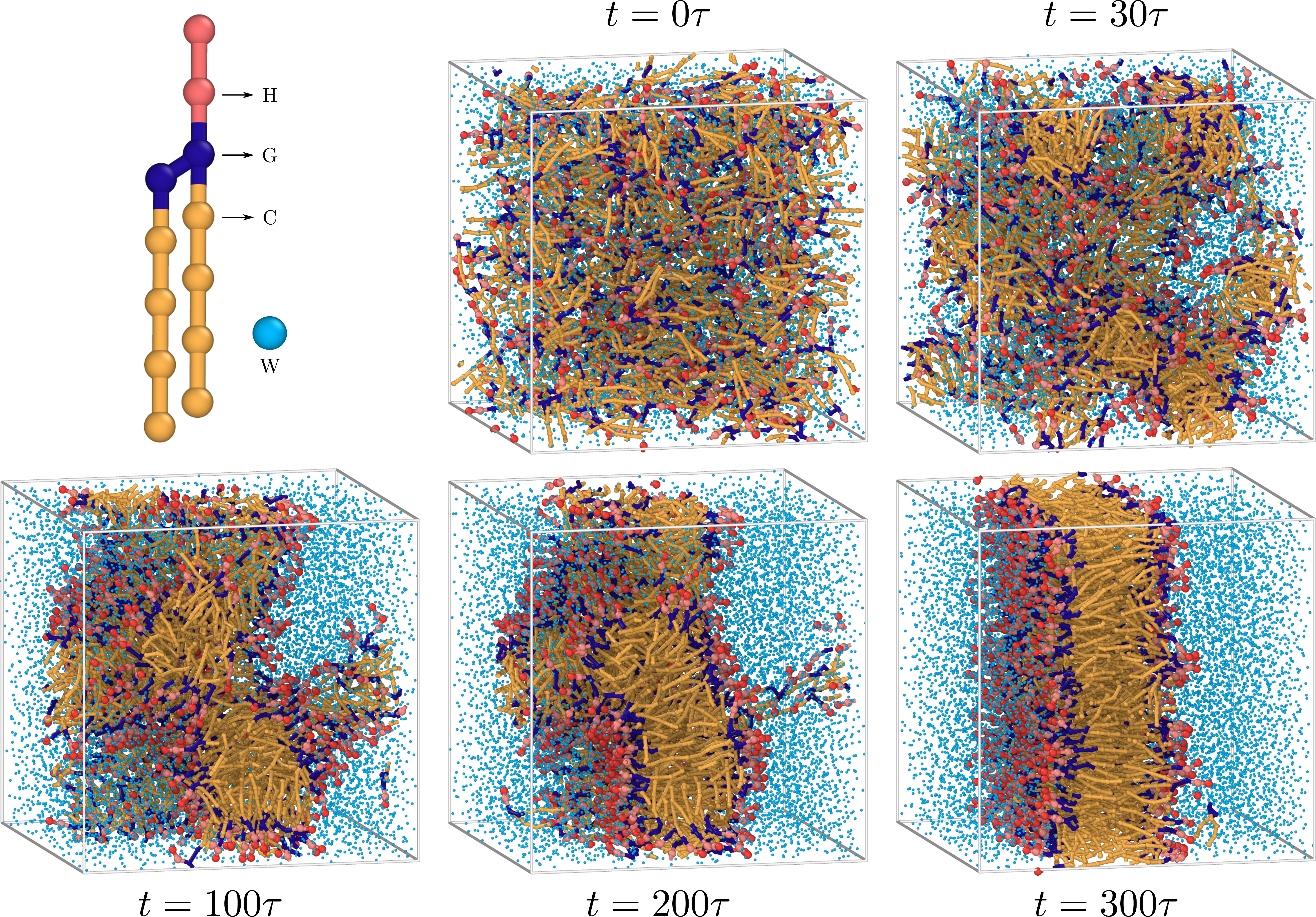}
\caption{\label{fig:2} 
MDPD CG model for DPPC lipid and water molecule. Snapshots illustrate the gradual formation of the DPPC bilayer starting from a random distribution 
of 512 lipid molecules that aggregate into clusters which then coalesce to 
produce the lipid bilayer. Time $\tau$ is in MDPD reduced units.
}
\end{figure}

% \begin{figure}[bt!]
% \centering
% \includegraphics[width=0.5\columnwidth]{FIGS/fig1-h.pdf}
% % \includegraphics[width=0.2\columnwidth]{FIGS/fig1-v.pdf}
% \caption{\label{fig:2} 
% MDPD DPPC coarse-grained model\cite{Zhu2021}.
% ``W'' represents three water molecules,
% ``C'' an alkane chain with four heavy atoms,
% ``G'' glycerol ester moiety,
% ``H'' the phosphate and choline headgroup.
% }
% \end{figure}

\section{Results and Discussion}\label{results}

We initiated MDPD and MD MARTINI simulations by randomly distributing
lipid molecules and water beads (Figure~\ref{fig:2}) within the simulation box.
The system was let to evolve until a defect-free bilayer structure 
emerged. Remarkably, the behavior of MD MARTINI lipids in our 
simulations closely mirrored results that have been previously 
reported \cite{Marrink2004}. In Figure \ref{fig:2}, 
we present the results of an MDPD simulation featuring 512 lipids. 
Initially, the lipids exhibited a tendency to aggregate into small 
clusters, which subsequently coalesced to form a more extensive
molecular assembly. Over time, this structure underwent further 
transformations, ultimately giving rise to a bilayer configuration.
Throughout these dynamic transitions, we observed the formation 
of transient defects, such as water pores and lipid bridges, 
analogous to the observations made in MD MARTINI simulations. 
These defects vanished over time, leading to the attainment of 
the final defect-free bilayer structure.
For comparison, Table \ref{tab:2} presents a time-based assessment
of MDPD and MD MARTINI simulations for various system sizes 
under consideration. Notably, our findings indicate that MDPD 
is capable of generating defect-free bilayers akin to
those produced by MD MARTINI simulations. However, a striking 
acceleration in computational efficiency was achieved through 
the utilization of soft, short-range interactions in MDPD.

\begin{table}[b]
\begin{tabular}{llll}
 System size & 128   & 256     & 512    \\ 
 \hline
  MDPD       &  20s  &  1.5min &  5min     \\ 
  MD     &  70s  &  5min   &  35min    \\
\end{tabular}
\caption{Time comparison of simulation runs on eight cores Intel Core i7-10700 @ 2.9Ghz for 
the formation of a bilayer with no defects. Successful formation highly depends on 
initial configuration but MDPD is always four to seven times faster than MD. 
Three system sizes were considered, with 128, 256 and 512 lipid molecules; 
all with a 10:1 proportion of water beads for both MDPD and MD.}
\label{tab:2} 
\end{table}

The area per lipid (APL) quantifies the surface area occupied 
by an individual lipid molecule. This measure is subject to 
influence by various factors, including the lipid type,
temperature, and the presence of other molecules within the system.
In our investigations, we adopted a method wherein the area
of plane aligned tangentially to the bilayer surface 
in the simulation box, was divided by half the 
total number of lipid molecules in order to determine the APL.
On the one hand, for the MDPD DPPC model, 
the resulting APL was found to be
$47.9{\AA}^2$, a value that aligns closely with experimental data 
and recent MD studies \cite{mot2023}. In the MD MARTINI model, 
on the other hand, the calculated APL was $61.9{\AA}^2$, 
a measure comparable to that of a DPPC bilayer 
at a temperature of $323K$ \cite{nagle2000}.

\begin{figure*}[bt!]
\centering
\includegraphics[width=\textwidth]{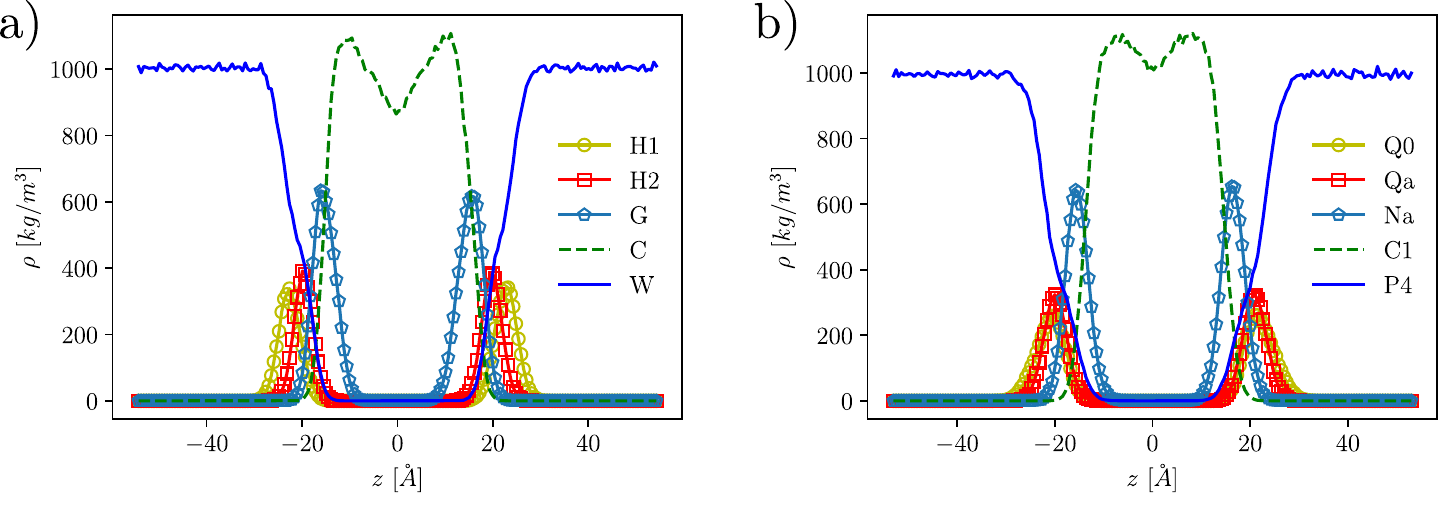}
\caption{\label{fig:3} 
 Comparison of Bilayer Thickness in MDPD and MD Simulations. (a) Density profile from MDPD simulation, with a bilayer thickness measured between the peaks of the top H bead (H1) equal to 45 Å. (b) Martini simulation results showing a bilayer thickness of 42 Å. Both simulations consist of 512 lipid molecules.
}
\end{figure*}

Another fundamental property we assessed was the bilayer thickness,
a parameter essential for characterizing the structural aspects 
of lipid bilayers. This property was determined by identifying
the peaks in the density distribution profile of the 
choline headgroup bead (H1 in the MDPD model and $Q_0$ in the 
MARTINI model), as illustrated in Figure \ref{fig:3}. 
The thickness of the MDPD bilayer was computed to be $45{\AA}$,
while that of the MD bilayer was $42{\AA}$. 
Notably, both values closely mirror experimental
measurements \cite{drabik2020}, affirming the accuracy and 
reliability of our simulation results in capturing the 
essential structural characteristics of DPPC lipid bilayers.

\begin{figure}[bt!]
\centering
\includegraphics[width=\columnwidth]{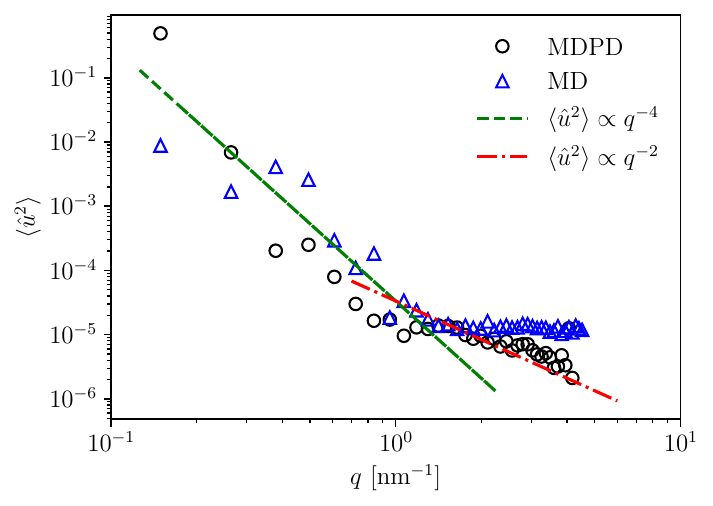}
\caption{\label{fig:4} 
 Spectrum intensities of the undulation modes on the bilayer. Long wavelengths obey the $q^{-4}$ behavior and transition around a $1$ nm wavelength to either a constant value in
 the MD simulations or to a $q^{-2}$ regime in MDPD simulation. The data is averaged for 100 time steps of a system consisting of 8192 lipids.
}
\end{figure}

Lipid bilayers, possessing elastic properties represented by the bending modulus, $k_c$, play a crucial role in membrane dynamics. The bending modulus quantifies the intensity of undulations within the lipid membrane and can be derived from the fluctuation spectrum \cite{tarazona2013}. In our analysis, equilibrated bilayer systems, each comprising 512 lipids, are replicated four times in both $x$ and $y$ directions, resulting in a large membrane consisting of 8192 lipids. Undulations are described as a surface function 
$u(x,y)$, representing the average position of two monolayers defined by the positions of phosphate head beads. The undulation spectrum $\hat{u} $is obtained by performing a Fourier transform on the surface function using the expression:
\begin{equation}
    \hat{u}(\mathbf{q}) = \frac{1}{N}\int u(\mathbf{x})e^{-i\mathbf{q}\cdot \mathbf{x}} \mathbf{dx}.
    \label{eq:fourier}
\end{equation} 
From a continuum model of the bilayer, the undulation intensities in Fourier space are
given by
\begin{equation}
    \langle \hat{u}^2 \rangle = \frac{k_B T}{A} \left( 
    \frac{1}{k_cq^4} + \frac{1}{\gamma q^2} \right)
    \label{eq:bendingModulus}
\end{equation}
where $A = L_xL_y$ is the lipid membrane area, $q = |\mathbf{q}| = \sqrt{q_x^2 + q_y^2}$ the wavevector length in the radial direction and $\gamma$ is surface tension. The analysis of the undulation spectrum, as illustrated in
Figure \ref{fig:4}, reveals that both MD and MDPD simulations exhibit the expected  
$q^{-4}$ scaling for the long wavelength modes. The dashed line represents a reference 
from the experimental bending modulus value of $k_c=5\times 10^{-20}$ J \cite{rawicz2000}. 
For wavelengths around $1$ nm and below, a transition occurs, with MD simulations showing a constant intensity and MDPD scaling exhibiting a $q^{-2}$ regime. This deviation may 
stem from differing surface tension conditions between simulations, with MDPD performed on 
an NVT ensemble and MD on an NPT ensemble employing a Nos\'e-Hoover barostat to maintain a 
zero surface tension condition. The  $q^{-2}$ line shown is the figure is obtained by 
considering $\gamma = 50$ mN/m. The difference could also be due to a possible protrusion 
regime present in the MDPD model or tilt angle of the lipid molecules as both can give 
rise to a $q^{-2}$ dominant regime \cite{brandt2011,may2007}.

\begin{figure}[bt!]
\centering
\includegraphics[width=\columnwidth]{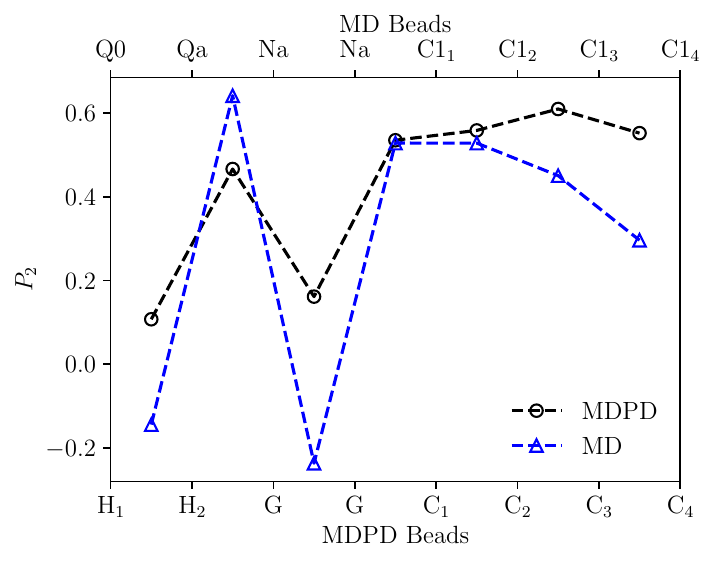}
\caption{\label{fig:5} 
 Second rank order parameter calculated for every bond in the lipid molecule. The data is 
 averaged for 100 time steps and over all the bonds in the bilayer consisting of 8192 lipids.
}
\end{figure}

To further compare the MD and MDPD methods, we compute the second-rank 
order parameter $P_2$ defined in Eq. \ref{eq:orderParameter} \cite{Lafrance1995,Chremos2016}. This is an 
important parameter that allows us to identify the alignment of the lipid molecules
with respect to the bilayer by measuring the angle $\theta $ between each bond and the
bilayer normal direction. 
\begin{equation}
    P_2 = \frac{1}{2} \left( 3\langle \cos^2\theta \rangle -1 \right)
    \label{eq:orderParameter}
\end{equation}
Figure \ref{fig:5} shows the $P_2$ value for each bond in the lipid molecule. A value of 
$P_2=1$ indicates perfect alignment, $P_2=0$ a random orientation and $P_2=-0.5$ 
anti-alignment. Our measurements suggest that the MDPD method results in lipids more 
aligned to the bilayer normal direction, possibly influenced by the presence of surface 
tension in the MDPD simulation \cite{Zhu2021}. Both methods exhibit a similar trend along 
the molecule chain, highlighting the robustness of the observed alignment characteristics.

\section{Conclusions}
\label{conclusions}

We have examined the efficacy of our proposed MDPD force-field in comparison with the widely recognized MARTINI force-field for CG MD 
simulations. Both methodologies demonstrated the capability to produce
defect-free lipid bilayer structures, accompanied by similar self-aggregation dynamics.
Furthermore, our investigation delved into the assessment of fundamental structural 
attributes of these lipid bilayers, specifically the determination of area per lipid, bilayer thickness, bending modulus, and orientation order parameter. Remarkably, both MDPD and MD MARTINI simulations yielded results
that closely approximated experimental observations.
A noteworthy advantage of the MDPD method lies in its utilization of soft, short-range interactions,
a feature that enables the realisation of faster simulations without
comprising on key properties of the system selected here for 
demonstration purposes.
This enhancement in computational efficiency holds paramount significance
in the context of large-scale molecular simulations,
potentially facilitating the exploration of increasingly intricate
and large molecular systems.

Not only our study highlights the effectiveness of the MARTINI "Lego" approach
when applied to the MDPD method, allowing for the representation of complex molecular
interactions in a CG fashion through bead parameterization
based on hydrophobic/hydrophilic characteristics, 
but also underscores the versatile nature of the MDPD force-field.
This adaptability paves the way for the incorporation of additional interaction levels,
diverse bead types, and varying bead sizes,
thereby extending its applicability to a wider spectrum of molecules and systems.
In conclusion, the MDPD approach emerges as a versatile and computationally 
efficient tool for the simulation of large-scale molecular systems,
offering an intriguing alternative to the MD MARTINI framework 
and holding promise for further advancements 
in CG simulations.

\section*{Supplementary Information}
Movie.mp4: Movie illustrates the formation of the DPPC lipid bilayer by using MDPD simulation.

\section*{Acknowledgments}
This research has been supported by the National 
Science Centre, Poland, under
grant No.\ 2019/34/E/ST3/00232. 
We gratefully acknowledge Polish high-performance 
computing infrastructure PLGrid (HPC Centers: ACK Cyfronet AGH) 
for providing computer facilities and support 
within computational grant no. PLG/2022/015747.

\section*{Data Availability Statement}
Data sets generated during the current study are available from the corresponding author on reasonable request.

\bibliographystyle{spphys}       % APS-like style for physics
\bibliography{bib}   % name your BibTeX data base

\end{document}